\documentclass{npw}
\usepackage{graphicx}

\def\d{\hbox{d}}
\def\be{\begin{equation}}
\def\ee{\end{equation}}
\def\bea{\begin{eqnarray}}
\def\eea{\end{eqnarray}}
\def\l{\label}
\def\r{{\bf r}}
\def\p{{\bf p}}

\def\om{\omega}

\def\hahat{\hat{H}}

\def\hahat0{\hat{H}_0}

\def\exp{\hbox{exp}}

\def\Im{{\mbox {\rm Im}}}

\def\d{\hbox{d}}

\def\e{e}

\def\siml{\hbox{\kern.1em \lower.6ex \hbox{$\sim$} \kern-1.12em
 \raise.6ex \hbox{$<$} \kern.1em}}
\def\simg{\hbox{\kern.1em \lower.6ex \hbox{$\sim$} \kern-1.12em
 \raise.6ex \hbox{$>$} \kern.1em}}

\begin{document}

\title{Isovector dipole-resonance structure within the effective surface
approximation}
\author{
  J P Blocki \\
  {\it National Centre of Nuclear Research, PL-00681 Warsaw, Poland}\\
  A G Magner\email{magner@kinr.kiev.ua} \\
  {\it Institute for Nuclear Research,  Kyiv 03680, Ukraine}\\
P Ring\\
{\it Technical Munich University,  D-85747 Garching, Germany}
}
\date{}
\maketitle

\begin{abstract}

The nuclear isovector-dipole strength structure is
analyzed in terms of the main
and satellite (pygmy) peaks within 
the Fermi-liquid droplet model.
Such a structure  is sensitive
to the value of the  surface symmetry-energy constant 
obtained analytically for different Skyrme forces
in the leptodermous effective surface approximation. 
Energies, sum rules and transition densities 
of the main and satellite peaks
for specific Skyrme forces are qualitatively 
in agreement with the experimental data
and other theoretical calculations.

\end{abstract}
\bigskip


 \section{Introduction}

The symmetry energy is a key quantity for studies of the fundamental 
properties of the exotic nuclei 
with a large excess of neutrons in the 
nuclear physics and astrophysics.
In spite of a very intensive study of
these properties, the surface symmetry energy 
constant is still rather not well determined
in the liquid droplet model (LDM) calculations 
\cite{myswann69}
and in more microscopic local density approach (LDA). 
In particular, in 
the extended Thomas-Fermi (ETF) approximation \cite{brguehak},  
or in models based on the Hartree-Fock (HF) method,
both with the Skyrme forces \cite{chaban,reinhardSV,pastore}, 
such a situation is rather in contrast to 
the well known volume symmetry energy constant.
The Skyrme force parameters responsible for the surface symmetry energy 
constant were obtained by comparing the theoretical calculations of the
basic static and dynamic characteristics of nuclei with the
experimental data. However, besides of  
difficulties in studying these characteristics away from the
nuclear stability line like subtracting 
curvature and shell effects from the total energy, they are basically 
insensitive to this constant.  Therefore, the 
experimental \cite{adrich,wieland,kievPygmy} and theoretical 
\cite{vretenar,ponomarev,nester1,nester2} investigations
of the finer structures like 
the so called  \cite{adrich,vretenar,ponomarev} pygmy 
resonances of the Isovector Dipole Resonance (IVDR) strength 
become especially interesting as being more sensitive to the value 
of the 
surface symmetry-energy constant. Simple and accurate enough 
analytical theories deriving this constant in terms of the 
Skyrme force parameters by
using the effective surface (ES) 
approximation are needed \cite{BMRV}.
This approximation exploits the property of saturation of the nuclear matter 
and a narrow diffuse-edge region in finite heavy nuclei. 
The ES is  defined as the location of points of the maximum  
density gradient. 
The orthogonal coordinate system related locally to the ES is specified by
a distance from a 
given point to the surface  and a tangent coordinate 
at the ES. In these coordinates, 
the variational condition of the nuclear energy minimum
 at the fixed particle and neutron-excess numbers within
the LDA is simplified significantly when using the leptodermous expansion 
over a small parameter $a/R \sim A^{-1/3} \ll 1$  in the LDM \cite{myswann69} 
or ETF approach \cite{brguehak,BMRV}  
($a$ is of the order of 
the diffuse edge thickness of the nucleus,  
$R$ is the mean curvature radius of the ES, and 
$A$ the number of nucleons in heavy nuclei). 
The accuracy of the ES  approximation in the ETF approach
with the spin-orbit (SO) and asymmetry terms
was checked  
by comparing results of the HF and ETF theories
for several Skyrme forces \cite{BMRV}.
The surface symmetry energy constant 
is sensitive to the
choice of the Skyrme force parameters in the corresponding gradient
terms of the symmetry energy density \cite{BMRV}.

In the present work, the  
surface symmetry-energy constant \cite{BMRV} as a function of the  
Skyrme force parameters is applied
to analytical calculations of the energies, energy weighted sum rules 
(EWSR) and transition densities for the main  
and its satellite (pygmy) peaks in the 
IVDR strength function within the 
Fermi-liquid droplet model (FLDM)
\cite{kolmagsh,belyaev}. We shall consider
the rare (zero sound) quasiparticles'-collision regime 
(close to the mean field approach),
in contrast to the opposite  frequent 
collision (hydrodynamic) regime of the Steinwedel-Jensen (SJ)
and Goldhaber-Teller models.

\vspace{-0.33cm}
\section{The Fermi-liquid droplet model}
\l{fldm}

For IVDR calculations,  the 
FLDM based on the linearized
Landau-Vlasov equations for the isoscalar $ \delta f^{}_{+}(\r,\p,t)$ 
and isovector $ \delta f^{}_{-}(\r,\p,t)$ distribution functions 
can be used in the phase space 
\cite{kolmagsh,belyaev},
\begin{equation}
{\partial \delta f^{}_{\pm} \over {\partial t}} 
\!+\! 
    {\p \over {m_\pm^*}}
   {\bf \nabla}_r  \left[ \delta f^{}_{\pm} 
\!+\! \delta \left(\e-\e^{}_F\right)
    \left(\delta \e_{\pm}
    \!+\! V_{\rm ext}^{\pm}\right)\right]
    \!=\!\delta St^{}_{\pm}.
\label{LVeq}
\end{equation}
Here $\e=p^2/(2m_\pm^*)$ is the equilibrium quasiparticle energy ($p=|\p|$)
and $\e_{{}_{\! F}}=(p_F^{\pm})^2/(2m_\pm^*)$ is the Fermi energy. 
The isotopic dependence of the Fermi momenta
$p_F^{\pm}=p^{}_F\left(1 \mp \Delta\right)~$
is given by a small parameter
$\Delta=2\left(1+F_{0}'\right)\;I/3~$, where 
$I=(N-Z)/A$ is the nuclear asymmetry parameter, $N$ and $Z$ are 
respectively the neutron and proton numbers
($A=N+Z$) \cite{kolmagsh}. The reason of having $\Delta$
is the difference between 
the neutron and proton potential depths due to the Coulomb interaction.
 The isotropic isoscalar $F_0$  and isovector 
 $F_0'$ Landau interaction constants are related 
to the incompressibility $K=6\e_{{}_{\! F}}(1+F_0) \approx 220-260$ MeV 
and the volume symmetry energy $J=2\e_{{}_{\! F}}(1+F_0')/3 \approx 30$ MeV
constants
of the nuclear matter, respectively.
The effective masses
$m_{+}^*=m(1+F_1/3)$ and $m_{-}^*=m(1+F_1^\prime/3)$ are determined in terms of 
the nucleon mass $m$ by anisotropic Landau constants $F_1$ and 
$F_1^\prime$.  Equations (\ref{LVeq}) are coupled
by  the dynamical variation 
of the quasiparticles' interaction  $ \delta \e_{\pm}$
with respect to the equilibrium 
value $p^2/(2m_\pm^*)$. The   time-dependent external field
$V_{\rm ext}^{\pm} \propto \exp(-i \omega t)$
is periodic with a frequency $\omega$. For simplicity,
the collision term $\delta St_{\pm}$ is calculated 
within the relaxation time $\tau(\om)$ approximation
accounting for the retardation effects 
due to the energy-dependent self-energy 
beyond the mean field approach \cite{kolmagsh,belyaev}.

Solutions of equations (\ref{LVeq}) are related to the dynamic multipole 
particle-density variations, $\delta \rho^{}_{\pm}(\r,t) \propto
Y_{L0}(\hat{r})$, where $Y_{L0}(\hat{r})$ are the spherical harmonics,
$\hat{r}=\r/r$.
These solutions can be found in terms of the
superposition of the plane waves over the angle 
of a wave vector 
${\bf q}$. Their time-dependence is periodic as the external field
$V_{\rm ext}^{\pm}$ is also periodic with the same frequency
$\om=p_F^\pm s^\pm q/m_{\pm}^*$ where
$s^{+}=s$, and $s^{-}=s \left(NZ/A^2\right)^{1/2}$. 
The factor $\left(NZ/A^2\right)^{1/2}$ accounts for conserving 
the position of the mass center 
for the isovector vibrations. 
The sound velocities $s^{}_n$ can be found from the dispersion equations
\cite{kolmagsh} as functions of 
the Landau interaction constants and 
$\omega \tau$. The ``out-of-phase'' particle-density vibrations
of the $s_1$ mode involve the ``in-phase'' $s_2$ ones inside the nucleus due
to the symmetry interaction coupling.

For small isovector and isoscalar multipole 
ES-radius vibrations of the finite neutron and proton 
Fermi-liquid drops around the
spherical nuclear shape, one has
$\delta R_{\pm}(t) = R \alpha_S^{\pm}(t) Y_{L0} ({\hat r})\;$ with
%
 a small time-dependent amplitudes
$\alpha_S^{\pm}(t) = \alpha_S^\pm \exp(-i \omega t)$. 
The macroscopic boundary conditions  
(surface continuity and force-equilibrium equations) at the ES  are given by
\cite{BMRV,kolmagsh,belyaev}:
\begin{equation}
u_{r}^{\pm}\Big|_{r=R} \!=\!
R \dot{\alpha}_S^{\pm} Y_{L0}({\hat r}),\quad
\delta \Pi_{rr}^{\pm}\Big|_{r=R}\!=\!
\alpha_S^\pm \overline{P}_S^\pm\;Y_{L0}({\hat r}).
\label{bound2}
\end{equation}
The left hand sides of these equations 
are the radial components
of the mean velocity field ${\bf u}={\bf j}/m$ 
and the momentum flux tensor $\delta \Pi_{\nu\mu}$ \cite{kolmagsh,belyaev}.
Their right hand sides are the ES velocities and capillary pressures. 
These pressures are proportional to the isoscalar and isovector 
surface energy constants $b_S^{\pm}$ \cite{BMRV},
\begin{equation}
\overline{P}_S^{\pm}=\frac23 
b_S^{\pm}\rho_{\infty} {\cal P}_{\pm}  A^{\mp1/3},\quad
b_S^{\pm} \propto {\cal C}_{\pm}\int_0^{\infty} \d r 
\left(\frac{\d \rho_{\pm}}{\d r}\right)^2,
\label{pressuresurf}
\end{equation}
where ${\cal P}_{+}= (L-1)(L+2)/2~$, ${\cal P}_{-}= 1~$, 
$\rho_{\pm}=\rho_n \pm \rho_p$ and  
$\rho_{\infty}=3/(4 \pi r_0^3)\approx 0.16$ fm$^{-3}$ is the 
density of the infinite nuclear matter. 
Coefficients $b_S^{\pm}$ are essentially determined by constants 
${\cal C}_{\pm}$
of the energy density in front of its gradient density terms 
$\propto \left(\nabla \rho_{\pm}\right)^2$.
The conservation of the mass center was taken into
account in derivations of  the second boundary conditions 
(\ref{bound2}) \cite{kolmagsh,belyaev}. Therefore, 
one has a dynamical equilibrium of forces acting at the ES. 

\section{Response and transition density}

The response function,
$\chi_{\pm}(\om)$,
is defined as a linear reaction of the average value of a
 single-particle operator $\hat{F}(\r)$  
in the Fourier $\omega$-representation to the external field.
For convenience, we may consider this field in terms of a similar 
superposition of the plane waves as  
for the distribution function
$\delta f_{\pm}$ \cite{kolmagsh,belyaev}.
 In the following, we will consider the 
long wave-length limit with
${\cal V}_{\rm ext}^{\pm}(\r,t)=\lambda_{\rm ext}^{\pm,\om}(t) {\hat F}(\r)~$ and
$\lambda_{\rm ext}^{\pm,\om}(t)=\lambda_{\rm ext}^{\pm,\om}~e^{-i(\om+i\eta)t}\;$,
where $\lambda_{\rm ext}^{\pm,\om}$ is the amplitude and $\omega$ is the frequency
of the external field ($\eta=+0$).
In this limit, the one-body operator  $\hat{F}(\r)$ becomes the standard 
multipole
operator, $\hat{F}(\r)=r^LY_{L0}(\hat{r})$ for $L\geq 1$.
The response function $\chi_{\pm}(\om)$ is expressed through the 
Fourier transform of the transition density 
$\rho_{\pm}^{\om}(\r)$ as
\begin{equation}
\chi_{\pm}(\om)=
-\int {\rm d}\r\;\hat{F}(\r)\;
\rho_{\pm}^{\om}(\r)/\lambda_{\rm ext}^{\pm,\om}.
\label{chicollrho}
\end{equation}
The transition density $\rho_{\pm}^{\om}(\r)$ is obtained
through the 
dynamical part of the particle density  $\delta \rho_{\pm}(\r,t)$ in  
macroscopic models 
in terms of the solutions $\delta f_\pm(\r,\p,t)$ 
of the Landau-Vlasov equations 
(\ref{LVeq}) with the boundary conditions (\ref{bound2}) 
as the same superpositions
of plane waves  
\cite{kolmagsh}:
$\delta \rho_{-}(\r,t)\!=\!\rho_{\infty} \alpha^{-}_S
\rho_{-}^{\om}(x) \;Y_{10}(\hat{r})\;e^{-i \om t},$
where
\be\l{drhom}
\rho_{-}^{\om}(x)=\frac{qR}{j_L'(qR)}\;
\left\{j^{}_1\left(\kappa\right)\;\rho(x) +
\frac{\d \rho_{-}}{\d x}\frac{g_{{}_{\! V}}}{g_{{}_{\! S}}}\;
\right\},
\ee
\be\l{gv}
g_{{}_{\! V}}=\int_0^{\overline{\rho}_0} 
\d \rho\; 
\sqrt{\rho(1+\beta \rho)}\;\kappa^3 j_{{}_{\! 1}}(\kappa)/(1-\rho),
\ee
\be\l{gs}
g_{{}_{\! S}}=\int_0^{\overline{\rho}_0} \d \rho\; 
\kappa^3\left[1 + {\cal O}({\widetilde{\rho}}^2)
\right],\,\, \kappa=\kappa_o\left[1+a x(\rho)/R\right], 
\ee
$\kappa_o=qR$. 
The first term in (\ref{drhom}) which is proportional
to the isoscalar dimensionless density $\rho=\rho_{+}$ 
(in units of $\rho_{\infty}$) 
accounts for 
the volume density vibrations.
The second term $\propto d\rho_{-}/dx$ where 
$\rho_{-}$
is a dimensionless isovector density $\rho_{-}$ (in units of $I\rho_{\infty}$)
corresponds to the density variations due to 
the shift of the ES. The particle number
and mass center position are conserved, and 
$j_n(\kappa)$ and $j_n'(\kappa)$ are the spherical Bessel functions
and their derivatives.
The upper integration limit $\overline{\rho}_{{}_{\! 0}}$ in (\ref{gv}) 
and (\ref{gs}) is defined
as the root of a transcendent equation $x(\rho)+R/a=0$.  
We introduced also the dimensionless SO
interaction parameter $\beta={\cal D}_{+} \rho_{\infty}/{\cal C}_{+}$
(${\cal D}_{+}=-9mW_0/16 \hbar^2$, where 
$W_0\approx 100-130$ MeV$\cdot$ fm$^5$ \cite{brguehak,BMRV}). 
Several other quantities were also defined by
$\widetilde{\rho}=(1-\rho)/c_{\rm sym}~$,  
$c_{\rm sym}=
a \left[J/\left(\rho_{\infty}\vert{\cal C}_{-}\vert\right)\right]^{1/2}\approx
2-4~$,
$a=\left[{\cal C}_{+}\; \rho_{\infty}\; K/(30\; b_V^2)\right]^{1/2}
\approx 0.5 - 0.6~$ fm is the 
diffuseness parameter, $b_{{}_{\! V}} \approx 16$ MeV. 
Simple approximate expressions for constants 
$b_S^{\pm}$ can be easily
derived in terms of
the elementary functions \cite{BMRV}.
Note that in these derivations we neglected curvature terms 
and shell corrections being of the same order. 
The isovector energy terms were obtained within the ES 
approximation with high accuracy up to the product of two
small quantities, $I^2$ and $(a/R)^2$.
The isovector equilibrium particle density $\rho_{-}(x)$
in (\ref{drhom}) can be given 
through the isoscalar one,  
$\rho_{+}(x) \equiv \rho(x)$. 
As shown \cite{BMRV}, 
the SO dependent density $\rho_{-}(x)$ 
is of the same order as $\rho(x)$.
The dependence of the 
isovector $\rho_{-}(x)$ on
different Skyrme force parameters, mainly ${\cal C}_{-}$ and $\beta$, is
the main reason of different values of the neutron skin.

With the help of the  
boundary conditions (\ref{bound2}), one can derive the response 
function (\ref{chicollrho}) \cite{kolmagsh},
\begin{equation}
\chi_{{}_{\! L}}(\om)\!=\!\sum_n \chi_{{}_{\! L}}^{(n)}(\omega)\!=\!
\sum_n {{\cal A}_{L}^{(n)}(\kappa_o)/ D_{L}^{(n)}(\om\!-\!i\Gamma/2)},
\label{respfuni}
\end{equation}
with $\om=p^{}_Fs_n \kappa_o \left(NZ/A^2\right)^{1/2}/(m^*R)~$
($m_{-}^*\approx m_{+}^*=m^*$).
This response function describes two modes, the main ($n=1$)
IVDR and its satellite  ($n=2$) 
as related to the out-of-phase $s_1$ and in-phase $s_2$ sound velocities,
respectively. We assume here that the ``main'' 
peak exhausts mostly
the energy weighted sum rule (EWSR) and the ``satellite'' corresponds
to a much smaller value of the EWSR. This two-peak structure is due to 
the coupling of the
isovector and isoscalar density-volume vibrations due to
 the neutron and proton 
quasiparticle interaction in (\ref{LVeq}).
The lowest poles ($n=1,2$) of the response function (\ref{respfuni}) 
are determined by the secular equation,
\begin{equation}
D_{L}^{(n)} \!\equiv\! j_L'(\kappa_o)\!-\!
\frac{3 \e_{{}_{\! F}}\kappa_oc_1^{(n)}}{2 b_S^{-} A^{1/3}}
\left[j_L(\kappa_o)\!+\!c_2^{(n)}j_L''(\kappa_o)\right]\!=\!0.
\label{seculeq}
\end{equation}
The width of an IVDR peak $\Gamma$ in (\ref{respfuni}) as an imaginary part
of the pole originated from the integral collision term  
$\delta St_{\pm}$
of the Landau-Vlasov equation.
For amplitudes one has ${\cal A}_{L}^{(n)} \propto \Delta^{n-1}$.
The complete expressions for ${\cal A}_{L}^{(n)}$ and
$c_i^{(n)}$ are given in \cite{kolmagsh,belyaev}.
Assuming a smallness of $\Delta$, one may 
call the $n=2$ mode as a ``satellite'' of the ``main'' $n=1$ peak.
On the other hand, other
factors such as a  collisional relaxation  time, 
 the surface symmetry energy constant $b_{S}^{-}$, and the particle
number $A$ lead sometimes to a re-distribution 
of the EWSR values among these two IVDR peaks. 
From (\ref{seculeq}), one can find their splitting \cite{kolmagsh}: 
\begin{equation}
\hbar | \om^{(1)}-\om^{(2)}| \approx
\left(10 / 3\right)^{3/2} \e_F^2 F_0^\prime I^2 / (|b_{S}^{-}| A^{2/3}).
\label{splitting}
\end{equation}
This relationship is important for the prediction
of distances between the ``satellite'' and the ``main'' peaks, depending
on the surface symmetry energy constant $b_{S}^{-}$,  
particle number $A$ and interaction constant $F_0'$.

According to the time-dependent HF approach based on the Skyrme forces 
\cite{nester1,nester2}, 
the energy of the pygmy
resonances in the isovector and isoscalar channels coincide approximately. 
On the other hand, the energy of the main peak of the 
Isoscalar Dipole Resonance (ISDR) is much larger 
than that of the IVDR. We may try to interpret within the 
FLDM the satellite
peak as some kind of the pygmy one, as observed experimentally, e.g., in some 
spherical  
isotopes $^{208}$Pb, $^{132}$Sn, and $^{68}$Ni 
\cite{adrich,wieland}. 
Therefore, we may calculate separately the 
neutron, $\rho_n^{\om}(x)$, and proton, $\rho_p^{\om}(x)$, transition 
densities for the satellite by calculating the isovector and isoscalar
transition densities at the same energy and in the same units as $\rho_{\pm}$,
$\rho^{\om}_{{}_{\! n \atop{p}}}(x)=
\left[\rho_{+}^{\om}(x) \pm \rho_{-}^{\om}(x)\right]/2$.

\begin{figure*}
\begin{center}
\includegraphics[width=0.65\textwidth]{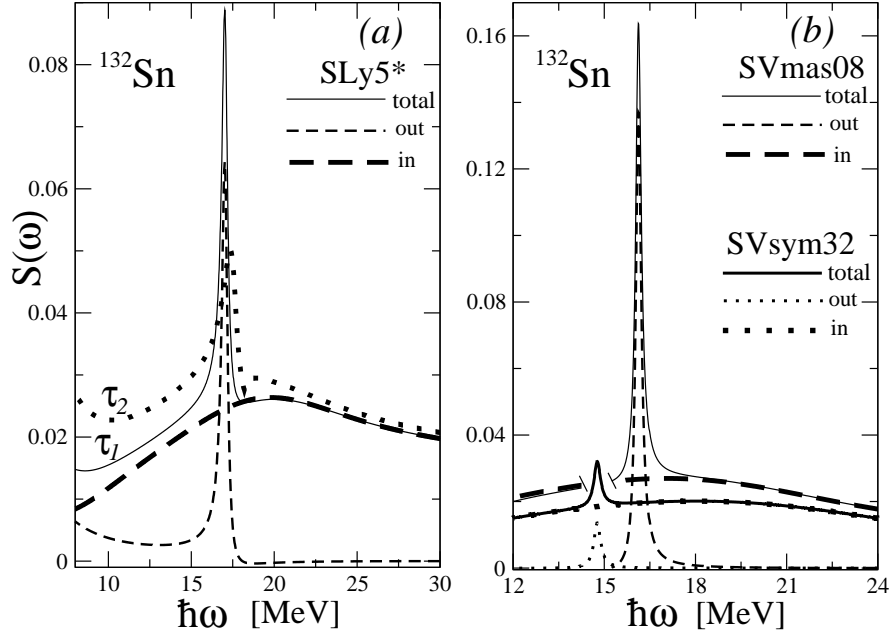}
\end{center}

\vspace{-0.2cm}
\caption{IVDR strength functions 
$S^{(n)}(\omega)$ 
vs the excitation
energy $\hbar \omega$ are shown for vibrations of the nucleus
$^{132}$Sn. {\it (a)}: The Skyrme force SLy5$^*$; frequent 
($n=1$ ``out''), and rare  ($n=2$ ``in'') dashed
curves and their 
total sum (solid) are shown;
[$\tau_1=4.4$ and $\tau_2=1.4 \cdot 10^{-21}$ s (dots)]. 
{\it (b)}: The same for two 
other Skyrme forces at the relaxation time
with the same frequency dependence as in $\tau_1$  
\cite{belyaev};
}
\label{fig1}
\end{figure*}

\vspace{-1.5cm}
\begin{figure*}
\begin{center}
\includegraphics[width=0.65\textwidth]{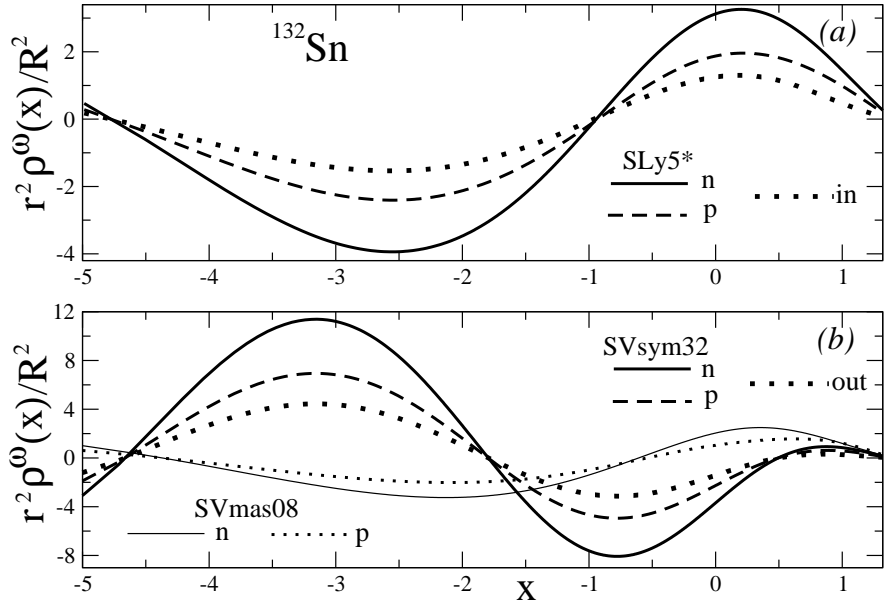}
\end{center}

\vspace{-0.2cm}
\caption{
The IVDR transition densities  $\rho_{-}^{\om}(x)$ 
(\ref{drhom}) multiplied by $(r/R)^2$
vs $x=(r-R)/a$ with 
{\it (a)} SLy5$^*$  \cite{pastore}
and {\it (b)} SVmas08 and SVsym32 \cite{reinhardSV}; 
rare dots show the ``in-phase'' ($n=2$) mode for SLy5$^*$ and
``out-of-phase'' ($n=1$) one for SVsym32
at the excitation energies $E_2$ 
(Table 1); the 
neutron $\rho_n^{\om}$ (n) and proton $\rho_p^{\om}$ (p) 
transition  densities  
for the energy $E_2$
are shown by the solid and dashed (frequent dots) lines for SLy5$^*$ and 
SVsym32 (SVmas08), respectively.}
\label{fig2}
\end{figure*}

\vspace{1.2cm}
\section{Discussion of the results}

The total IVDR strength function being 
the sum of the two (``out-of-phase'' $n=1$ and 
``in-phase'' $n=2$) 
strength functions 
[solid lines in 
Fig.\ \ref{fig1}]
has rather a remarkable shape asymmetry. 
In Fig.\ 1{\it (a)} for the SLy5$^*$ and {\it (b)} SVmas08 cases, 
one has the 
``in-phase'' satellite on right of the 
main ``out-of-phase''
peak. An enhancement on its left for SLy5$^*$ is due to
the increasing of the ``out-of-phase'' strength (frequent dashed) curve 
at too small energies. A more 
pronounced enhancement is seen for the SVsym32 force in 
Fig.\ \ref{fig1}{\it (b)} because of 
the ``out-of-phase'' satellite (frequent dots),  
that is in better accordance with the 
experimental data 
\cite{adrich,wieland}. This IVDR strength structure is 
in contrast to the SVmas08 result 
with the  dominating ``out-of-phase'' peak shown in {\it (b)}.
As seen from comparison in Fig.\ 1{\it (a)} with {\it (b)}, 
the shape structure of the 
total IVDR strength
function depends significantly on the surface symmetry 
energy coefficient $k_{{}_{\! S}}=b_{S}^{-}/I^2$ 
where $b_{S}^{-} \propto I^2$ \cite{BMRV}. 
Notice also that one has its strong dependence on the 
relaxation time $\tau$ 
[see, e.g., curves for the relaxation times $\tau_1$ and 
$\tau_2$ in Fig.\ 1{\it (a)}]. 
The ``in-phase'' strength component with  
rather a wide maximum is weakly dependent 
on the choice of the Skyrme forces \cite{chaban,reinhardSV,pastore}.

Fig.\ \ref{fig2} shows the transition particle densities 
$\rho^{\om}(x)$ 
(\ref{drhom}) as functions of the dimensionless 
radial variable $x=(r-R)/a$ for the isovector vibrations in $^{132}$Sn
at the same Skyrme forces as in Fig.\ 1.
Results of these calculations look qualitatively 
similar to those of \cite{vretenar,nester1,nester2}. 
The neutron particle density is always significantly 
larger than the proton one in the surface region $|x| \siml 1$. The 
difference is in a distance between the neutron transition 
density curve and the proton one, but
one has always the neutron skin around a symmetric core.

Within the ES approximation,  
the surface symmetry 
energy coefficient $k_{{}_{\! S}}$ 
for three Skyrme forces \cite{reinhardSV,pastore} with 
a two-peak structure
is shown in Table 1.
The isovector energy
coefficient $k_{{}_{\! S}}$  
is more sensitive to the choice
of the Skyrme forces than the isoscalar one $b_S^{+}$ 
\cite{BMRV}.
The magnitude of $k_{{}_{\! S}}$ for the most of SLy 
\cite{chaban}, SkI and SV \cite{reinhardSV} forces
is significantly larger than for other ones (the corresponding isovector 
stiffness $Q$ depending also on the neutron skin
\cite{myswann69} 
was obtained analytically \cite{BMRV}). 

\vspace{-0.1cm} 
According to (\ref{pressuresurf}),
the most responsible fitting parameters in the Skyrme HF approach
which result in significant differences in $k_{{}_{\! S}}$ 
(or $b_S^{-}$)  
values are the key constants ${\cal C}_{-}$ in
gradient terms of the energy density,
$k_{{}_{\! S}}\propto {\cal C}_{-}$ \cite{BMRV}.
The constant ${\cal C}_{-}$ is strongly 
dependent on different Skyrme 
forces (even in sign), in contrast to the isoscalar energy density
constants $ b_S^{+}$ \cite{BMRV}. 
There are still unclear interpretations
of the experimental results \cite{adrich,wieland}
which would determine $k_{{}_{\! S}}$  well enough as 
the mean Isovector Giant Dipole Resonance (IVGDR)
energies  are 
almost insensitive to $k_{{}_{\! S}}$ for
different Skyrme forces \cite{BMRV}. 
Another reason for so different $k_{{}_{\! S}}$ values might be related
to difficulties in extracting $k_{{}_{\! S}}$  
directly from HF calculations due to the curvature 
and quantum (shell) effects.
We have to go also away from the nuclear stability line to 
extract uniquely the coefficient $k_{{}_{\! S}}$ out of the dependence: 
$b_S^{-} \propto I^2=(N-Z)^2/A^2$.
For exotic nuclei, one has more problems to relate $k_{{}_{\! S}}$ to the 
experimental data with a good enough precision. 
We emphasize that there is an abnormal
behavior of the isovector surface constant $k_{{}_{\! S}}$  
as related to the constant 
$C_{-}$ of the energy density \cite{BMRV}.
This is in contrast to all
other Skyrme forces where $k_{{}_{\! S}}<0$ as ${\cal C}_{-} <0$ with a normal 
positive isovector
stiffness $Q$ of the stable neutron-skin vibrations.  
For specific Skyrme forces, such as RATP \cite{chaban},
SkI and SV \cite{reinhardSV}, $k_{{}_{\! S}}$ is positive 
as ${\cal C}_{-}> 0$, that could mean one has an abnormal negative isovector 
stiffness $Q$ of unstable neutron-skin vibrations. For some Skyrme forces
like SkT6 \cite{chaban}, the isovector stiffness even diverges,  $Q=\infty$,
because of $k_{{}_{\! S}}=0$ ($C_{-}=0$). Therefore, it is 
impossible to excite the isovector neutron-skin vibrations and,
as in the hydrodynamic SJ model, we cannot expect pygmy resonances 
in the IVDR strength.

The FLDM calculations of the IVDR energies
$E_n$ and EWSR $S_n$ in terms of the strength distributions
 $S^{(n)}(\omega)=\Im \chi_1^{(n)}(\omega)/\pi$ [see (\ref{respfuni})]
at $n=1$ and $n=2$ peaks $\omega=\omega_n$ 
[$S^{(n)}=S^{(n)}(\omega_n)$, Fig.\ 1] are shown in Table 1. 
The averaged constants
$D=(D_1 S^{(1)} + D_2 S^{(2)})/[S^{(1)} + S^{(2)}]$ with $D_n=E_n A^{1/3}$ 
for the neutron-rich spherical
(double magic) nuclei $^{208}$Pb, $^{132}$Sn, and $^{68}$Na are shown too.
We call the first one the ``main'' peak defined as it is 
exhausting mainly 
the EWSR,
$S_n=S^{(n)} E_n/[E_1 S^{(1)} + E_2 S^{(2)}]$, in contrast to the ``satellite''
one with a smaller EWSR contribution.
The IVDR energies
$E_n=\hbar \om_n=\hbar \kappa_o^{(n)} p_F^{-} s_{n}^{-} /(m^*R)$ were obtained by 
calculations of the wave numbers $\kappa_o^{(n)}$ as poles 
($\kappa_o^{(n)}=q_nR$)
of the response functions $\chi_n(\om)$
 (\ref{respfuni})  
for the sound velocities $s_n^{-}$ \cite{kolmagsh,belyaev}. 
The $n=1$ peak coming from
the out-of-phase volume vibration of 
the neutron-vs-proton particle densities
with a sound velocity $s_{{}_{\! 1}}$ and the $n=2$ peak  related to
the in-phase isoscalar-like sound with  a velocity 
$s_{{}_{\! 2}}$ are excited by the same isovector-like out-of phase
vibration of the neutron vs proton Fermi-liquid drop surfaces.  
Typically, these modes can be assigned respectively 
as main and satellite
peaks because the strength values  
$S^{(2)}(\omega) \propto
\Delta \propto I$ are smaller than $S^{(1)}(\omega)$ of the zero order in $I$. 

\vspace{-0.3cm}
\noindent
\hspace{1cm}
\begin{table*}[pt]
\begin{tabular}{|c|c|c|c|c|c|c|c|c|c|c|c|c|c|}
\hline
  
Skyrmes & $^{A}$X & $E_1$ & $S_1$ & $E_2$ & $S_2$ & $D^{}_{FLD}$ & $D^{}_{HD}$ 
& $k_S$ & $Q$ 
& ${\cal C}_{-}$ & $\beta$ & $\tau$ \\ 
& & MeV & \% & MeV & \% & MeV & MeV &  MeV &  MeV & MeV$\cdot$fm$^{5}$ & & 
$\times 10^{-21}$s\\
\hline
SLy5$^*$& $^{208}$Pb &  14.5  & 74 & 17.1  
  & 26 & 89 & 84  & -3.94 & 107 & -24.2 & -0.58 & 6.0 \\
&  $^{132}$Sn & 17.0  & 68 & 19.8
& 32 & 91 & 83 &  &  &  & & 4.4\\
&  $^{68}$Ni & 20.9 & 61 &  25.0
   & 39 & 91 & 82 &   &  & &  & 2.9 \\
SVmas08& $^{208}$Pb &  13.9  & 89 & 14.8  
  & 11 & 83 & 101  & 12.4 & -60 & 36.9 & -0.51 & 7.4 \\
&  $^{132}$Sn & 16.1  & 83 & 17.1
& 17 & 83 & 104 &  &  &  & & 5.5\\
&  $^{68}$Ni & 20.3 & 83 &  21.9
   & 17 & 84 & 110 &   &  & &  & 3.5 \\
SVsym32& $^{208}$Pb &  15.6  & 53 & 12.8  
  & 47 & 84 & 97  & 3.41 & -118 & 26.0 & -0.47 & 9.9 \\
&  $^{132}$Sn & 18.1  & 64 & 14.8
& 36 & 85 & 98 &  &  &  & & 7.4\\
&  $^{68}$Ni & 23.4 & 68 &  18.7
   & 32 & 88 & 101 &   &  & &  & 4.7 \\
\hline 
\end{tabular}

\vspace{-0.15cm}
\caption{ {\small Energies $ E_n$ and EWSR 
$S_n~$  
($S_{1}+S_{2}=100 $ \%)
for the out-of-phase $n=1$ and in-phase $n=2$ IVDR 
(or otherwise)
 are shown for the SLy5$^*$  \cite{pastore} and SVmas08 \cite{reinhardSV}
(but for SVsym32) Skyrme forces; $k_{{}_{\! S}}$, $Q$ \cite{BMRV}, 
${\cal C}_{-}$, $\beta$  
and $\tau$ \cite{kolmagsh} at the IVGDR peak
are explained in the text; the 7th and 8th columns are the mean
(IVGDR) energy constants  
$D^{}_{FLD}(A)$ 
averaged with the strength distributions
of the desired peaks 
and  $D^{}_{HD}(A)$ \cite{BMRV}
 calculated within the FLD and HD (SJ) models, respectively. }
}
\end{table*}

\vspace{-0.2cm} 
A satellite with a relatively smaller 
EWSR contribution can be
interpreted within the FLDM, as a
Pygmy Dipole Resonance 
\cite{vretenar,nester1,nester2}. 
These resonances were found for several nuclear isotopes
with the Skyrme forces SLy5$^*$ \cite{pastore} 
or SVmas08 and SVsym34 \cite{reinhardSV} 
[see, e.g., 
Fig.\ \ref{fig1}{\it (a)} or {\it (b)} for the nucleus $^{132}$Sn]. 
A smaller contribution at SLy5$^*$ and SVmas08
of the in-phase (isoscalar-like volume-compression) density vibrations to 
the main (out-of-phase) IVDR peak
is associated with those discussed in \cite{nester1,nester2} 
but on its right side. As seen from Fig.\  \ref{fig1}{\it (b)}, 
one has different IVDR strength structures  
mainly due to the difference in $k_s$: 
The in-phase SVsym32 mode becomes 
even slightly dominating.  
Note that our FLDM 
splitting is alternative to the quantum isotopic one 
which fails for heavy nuclei \cite{kolmagsh}.

\vspace{-0.2cm}
\section{Summary}
 
Expressions for the surface symmetry energy constant 
$k_{{}_{\! S}}$
derived from the simple isovector solutions
of the particle density and nuclear energy within the leading ES 
approximation \cite{BMRV} are used in calculations
of the energies, sum rules for the  IVDR strength and transition
densities
within the FLDM  \cite{kolmagsh,belyaev} for some  
Skyrme forces \cite{reinhardSV,pastore}.  
The constant $k_{{}_{\! S}}$  
depends much on the critical
parameters of the Skyrme forces, mainly through  
${\cal C}_{-}$  
of the  density gradient terms in the isovector part of the energy density
and SO interaction constant $\beta$.

IVDR strengths are split into the main
and satellite peaks.
The mean (IVGDR) energies and EWSR values are in 
a fairly good agreement with the 
experimental data. 
According to our results for the basic IVDR characteristics,  
the neutron and proton transition densities 
[Fig.\ \ref{fig2}],
we may interpret semiclassically the IVDR satellites as 
some kind of the pygmy resonances. 
Their energies, sum rules and n-p transition densities
obtained analytically within the semiclassical FLDM approximation 
are sensitive to the surface symmetry energy constant 
$k_{{}_{\! S}}$. Therefore, their comparison with the  
experimental
data can be used for the evaluation of $k_{{}_{\! S}}$.

As perspectives, it would be worth to
apply our results to the systematic IVDR calculations of the pygmy resonances
within the Fermi-liquid droplet model \cite{kolmagsh,belyaev} 
and the isovector low-lying collective states within the periodic orbit
theory \cite{gzhmagfed}. 
They all are expected to be more sensitive
to the values of $k_{{}_{\bf S}}$. Our analytical approach without any fitting
is helpful for further 
study of the 
effects in the surface symmetry energy. 

\vspace{-0.1cm}
\begin{ack}
Authors thank 
M. Kowal, T. Kozlowski,
V.O. Nesterenko, P.-G. Reinhard, and
J. Skalski 
for many useful discussions.  One of us (A.G.M.) is
also very grateful for a nice hospitality during his working visits at the
National Centre of Nuclear Research in Poland.
This work was partially supported by the Deutsche
Forschungsgemeinschaft Cluster of Excellence Origin and
Structure of the Universe (www.universe-cluster.de).
\end{ack}

\end{document}